\documentclass[aps,twocolumn,showpacs,showkeys]{revtex4}
\usepackage{amsmath}
\usepackage{amssymb}
\usepackage{graphicx}

\begin{document}

\title{Oscillation of the spin-currents of cold atoms on a ring due to
light-induced spin-orbit coupling}
 \author{Xie Wen-Fang (解文方)}
 \affiliation{School of Physics
and Electronic Engineering, Guangzhou University, Guangzhou
510006, People's Republic of China}
 \author{He Yan-Zhang (贺彦章)}
 \author{Bao Cheng-Guang (鲍诚光)}
 \thanks{Corresponding author: stsbcg@mail.sysu.edu.cn}
 \affiliation{School of
Physics and Engineering, Sun Yat-Sen University, Guangzhou
510275, People's Republic of China.}


\date{\today}

\begin{abstract}
The evolution of two-component cold atoms on a ring with
spin-orbit coupling has been studied analytically for the case
with N noninteracting particles. Then, the effect of
interaction is evaluated numerically via a two-body system. Two
cases are considered: (i) Starting from a ground state the
evolution is induced by a sudden change of the laser field, and
(ii) Starting from a superposition state. Oscillating
persistent spin-currents have been found. A set of formulae
have been derived to describe the period and amplitude of the
oscillation. Based on these formulae the oscillation can be
well controlled via adjusting the parameters of the laser
beams. In particular, it is predicted that movable stripes
might emerge on the ring.
\end{abstract}



\maketitle

\section{Introduction}

It is well known that the study of the motion of charged
particles under a magnetic field is an essential topic in both
macroscopic and microscopic physics. In particular, a number of
distinguished quantum mechanic phenomena, such as the
Aharonov-Bohm (A-B) oscillation and the fractional quantum Hall
effect (FQHE), are caused by the magnetic gauge
field.\cite{ya,dct,rbl} After the experimental realization of
the condensation of neutral atoms with nonzero spin,\cite{jst}
a great interest is to create a light-induced gauge vector
field to achieve the spin-orbit coupling (SOC) so that various
magnetic-electronic phenomena of charged particles occurring in
condense matters can be copied into the world of condensates of
neutral atoms.\cite{jda,hz1,vga,ngo,hz2} Since 2005 there are
proposals by the theorists for producing the light-induced
gauge vector field for the condensates with multi-component
neutral cold atoms.\cite{kos,jru,slz,xjl,gju,bma} The first
experimental realization of the SOC in condensates was
implemented in 2009 by dressing the cold atoms with two counter
propagating polarized laser beams.\cite{yjl1,yjl2} This
technique opens a new perspective in the field of BEC, more and
more experimental\cite{yjl3,yjl4,sb,jyz} and theoretical
\cite{yli,toz,zch} results are reported. It was found that the
total magnetization (or spin-polarizability, which measures the
difference of the densities of the spin-components) varies with
time. It implies the existence of
spin-currents.\cite{slz,yjl1,yli} Besides, the collective-mode
dynamics has been studied, the spatial motions along different
directions could be correlated (say, a dipole mode might induce
a breathing mode in its perpendicular plane).\cite{zch} The
instability caused by the collective mode has also been
studied.\cite{toz}

On the other hand it is now possible to trap a condensate in
ring geometry. Long-lived rotational superflows have been
induced in this kind of
systems.\cite{sb,sg,asa,cr,kh,ar,bes,sm,kcw,xue} The creation
of the spin-currents of two-component condensates on a ring,
due to the introduction of the laser beams, has been
experimentally realized.\cite{sb} It was found that the
stability depends strongly on the initial ratio of the two
components.\cite{sb} However, the oscillation of the
spin-currents on the ring has not yet been observed.

Obviously, the study of the cold spin-currents (i.e., the
current of each spin-component) caused by the SOC is just
beginning. It is expected that the oscillation of the
spin-currents might also emerge in the ring geometry (because a
circular motion is equivalent to a linear motion with a
periodic boundary condition). To confirm this suggestion, a
one-dimensional model of the two-component condensates on a
ring under SOC is adopted in this paper. The aim is to clarify
from the theoretical aspect how the details of oscillation (the
period and amplitude) depend on the parameters of the laser
beams. We believe that the interaction might not be crucial.
Therefore, the interaction is neglected firstly so as to
emphasize the decisive role of the laser beams. Then, the
effect of the interaction is evaluated via a two-body system.

\section{Time-dependent solution}

Let the quasi-spin $\hat{s}$ be introduced to describe the two
components of an atom as usual. The atomic state with $s_z=1/2$
$(-1/2)$ is named the up- (down-) state. When two
counter-propagating and polarized laser beams are applied, due
to the SOC, the two components of atoms might move towards
opposite directions along the beams and can transform to each
other via a spin-flip. When the interaction is neglected, the
Hamiltonian is just $H=\Sigma_i\hat{h}_i$, where $\hat{h}_i$ is
for the $i$-th particle. Neglecting the subscript $i$,
\begin{equation}
 \hat{h}
  =  \left(
     \begin{array}{ll}
      -\frac{\partial^2}{\partial\theta^2}+\delta , & \gamma e^{-2i\beta\theta } \\
       \gamma e^{2i\beta\theta },                   & -\frac{\partial^2}{\partial\theta^2}-\delta
     \end{array}
     \right),
\end{equation}
where $\theta$ is the azimuthal angle along the ring.\cite{hz2}
The unit of energy is hereafter $E_{\mathrm{unit}}
\equiv\hbar^2/(2mR^2)$, where $m$ is the mass of an atom, $R$
is the radius of the ring. $\beta\equiv k_0 R$, where $2k_0$ is
the momentum transfer caused by the two lasers. $\gamma
\equiv\frac{\Omega}{2E_{\mathrm{unit}}}$, where $\Omega/2$ is
the strength of Raman coupling causing the spin-flips
accompanied by the momentum transfer. Let
$\varepsilon_{\mathrm{split}}$ be the Zeeman energy difference
between the two spin-states, and $\omega_{\delta}$ is the
frequency difference between the two laser beams. We define the
Raman detuning $\delta \equiv(\varepsilon_{\mathrm{split}}
-\hbar\omega_{\delta})/(2E_{\mathrm{unit}})$. In addition to
$\gamma$, $\delta$ is an important quantity because it can be
tuned so that the related process could be energy-adapted (see
below). Note that, due to the ring geometry, $2\beta$ must be
an integer. It implies that the transfer of momentum will be
suppressed unless the transfer is close to a specific set of
values depending on $R$. Thus, an additional constraint is
required under the ring geometry.

To illustrate the connection with conventional SO coupling, we
define a $U$-transformation so that the up-state (down-state)
is multiplied by a factor $e^{i\beta\theta}$
($e^{-i\beta\theta}$). Then, by introducing the Pauli matrices
$\sigma_z$ and $\sigma_x$, $U\hat{h}U^{-1}$ can be written in a
more familiar form as \cite{hz2,wz,yl,slz,yjl1}
\begin{equation}
 U\hat{h}U^{-1}
  =  (-i\sigma_I\frac{\partial}{\partial\theta}
      -\beta\sigma_z)^2
    +\delta\sigma_z
    +\gamma\sigma_x,
\end{equation}
where $\sigma_I$ is just a unit matrix with rank 2.

$\hat{h}$ has two groups of eigenstates, they can be written as
\cite{wz,yl,yjl1}
\begin{eqnarray}
 & & \psi_k^{(+)}
  =  \sin(\rho_k)
     \varphi_{k-\beta,\uparrow}
    +\cos(\rho_k)
     \varphi_{k+\beta,\downarrow},  \\
 & & \psi_k^{(-)}
  =  \cos(\rho_k)
     \varphi_{k-\beta,\uparrow}
    -\sin(\rho_k)
     \varphi_{k+\beta,\downarrow},
\end{eqnarray}
where
\begin{eqnarray}
 \varphi_{k\pm\beta,\uparrow}
 &=& \frac{1}{\sqrt{2\pi}}
     e^{i(k\pm\beta)\theta}
     \binom{1}{0},  \\
 \varphi_{k\pm\beta ,\downarrow}
 &=& \frac{1}{\sqrt{2\pi}}
     e^{i(k\pm\beta)\theta}
     \binom{0}{1}.
\end{eqnarray}
$k$ is an integer (half-integer) when $\beta$ is an integer
(half-integer), $\sin\rho_k
=s_{\gamma}\sqrt{(a_k-2k\beta+\delta)/(2a_k)}$,
$\cos\rho_k=\sqrt{(a_k+2k\beta-\delta)/(2a_k)}$, $s_{\gamma}$
is the sign of $\gamma$,
$a_k=\sqrt{(2k\beta-\delta)^2+\gamma^2}$, and $\rho_k$ is
ranged from $-\pi/2\rightarrow\pi/2$. One can see that the
eigenstates have two notable features:

(i) Each of them contains both the up and down components, they
are propagating with angular momenta $k-\beta$ and $k+\beta$,
respectively. Where $2\beta$ is proportional to the momentum
transfer caused by the counter-propagating laser beams.

(ii) In each $\psi_k^{(\pm)}$ the weights of the two components
can be controlled by tuning $\delta$. When $\delta<2k\beta$, we
have $-\pi/4\leq\rho_k\leq\pi/4$, and therefore the weight of
the up-component is larger in $\psi_k^{(-)}$ but smaller in
$\psi_k^{(+)}$. Whereas when $\delta>2k\beta$, $\psi_k^{(-)}$
will contain more down-component. In particular, when $\delta$
is tuned so that $\delta=2k\beta$, the weights of the two
components in all the eigenstates are equal.

The eigenenergies of $\psi_k^{(\pm)}$ are
$E_k^{(\pm)}=k^2+\beta^2\pm a_k$. Obviously, $E_k^{(-)}\leq
E_k^{(+)}$.

The time-dependent solution $\psi(\theta,t)$ of the
single-particle Schr\"{o}dinger equation starting from an
initial state $\psi_{\mathrm{init}}$ can be formally written as
\begin{eqnarray}
 \psi(\theta,t)
 &=& e^{-i\tau\hat{h}}
     \psi_{\mathrm{init}} \nonumber \\
 &=& \sum_{k\lambda}|
     \psi_k^{(\lambda)}\rangle
     e^{-i\tau E_k^{(\lambda)}}\langle
     \psi_k^{(\lambda)}|
     \psi_{\mathrm{init}}\rangle,
\end{eqnarray}
where $\tau\equiv tE_{\mathrm{unit}}/\hbar$ (for $^{87}$Rb and
$R=12\mu m$ as given in \cite{sb}, $t=0.398\tau\sec$),
$\lambda=\pm$, $k$ runs over the (half-) integers when $\beta$
is an (half-) integer.

\section{Evolution caused by a sudden change of the laser field}

In the experiment reported in \cite{jyz} a strong oscillation
of the magnetization induced by a sudden change of the laser
field was reported. It is assumed that a set of parameters
$\beta$, $\gamma_1$, and $\delta_1$ (the first set) are given
initially, and the initial state is just the ground state
(g.s.) of $\hat{h}$ as $\psi_{\mathrm{init}}
=\psi_{\mathrm{gs}}
=\cos(\rho_q^{(1)})\varphi_{q-\beta,\uparrow}-\sin(\rho_q^{(1)})\varphi_{q+\beta,\downarrow}$,
where $q$ depends on the three parameters so that the
associated energy $E_q^{(-)}$ is the lowest. The superscript in
$\rho_q^{(1)}$ implies that it is obtained from the first set
of parameters. Then, $\gamma_1$ and $\delta_1$ are suddenly
changed to $\gamma_2$ and $\delta_2$ (the second set).
Accordingly, we have a new Hamiltonian and a new set of
eigenstates. With them $\psi(\theta,t)$ becomes
\begin{equation}
 \psi(\theta,t)
  =  e^{-i\tau(q^2+\beta^2)}
     ( f_u
       \varphi_{q-\beta,\uparrow}
      -f_d
       \varphi_{q+\beta,\downarrow}),
\end{equation}
where $f_u
=\cos(\rho_q^{(1)})\cos(a_q^{(2)}\tau)+i\cos(2\rho_q^{(2)}-\rho_q^{(1)})\sin(a_q^{(2)}\tau)$,
and $f_d
=\sin(\rho_q^{(1)})\cos(a_q^{(2)}\tau)+i\sin(2\rho_q^{(2)}-\rho_q^{(1)})\sin(a_q^{(2)}\tau)$.
Where $a_q^{(2)}$ and $\rho_q^{(2)} $ are obtained from
$\gamma_2$ and $\delta_2$. From $\psi(\theta,t)$, we know that
the two components of the g.s., $\varphi_{q-\beta,\uparrow}$
and $\varphi_{q-\beta,\uparrow}$, remain unchanged. However,
the coefficients of composition are no more constants but
oscillate with $\tau$.

Let the time-dependent densities of the up- and down-component
be defined from the identity $\psi^+(\theta,t)\psi(\theta,t)
\equiv n_{\uparrow}(\theta,\tau)+n_{\downarrow}(\theta,\tau)$.
Then, we have
\begin{eqnarray}
 n_{\uparrow}
 &=& \frac{1}{2\pi}
     \{ \cos^2(\rho_q^{(1)})
       +[ \cos^2(2\rho_q^{(2)}-\rho_q^{(1)})
         -\cos^2(\rho_q^{(1)}) ] \nonumber \\
 & &    \times
        \sin^2(a_q^{(2)}\tau) \}, \\
 n_{\downarrow}
 &=& \frac{1}{2\pi}
     \{ \sin^2(\rho_q^{(1)})
       +[ \sin^2(2\rho_q^{(2)}-\rho_q^{(1)})
         -\sin^2(\rho_q^{(1)}) ] \nonumber \\
 & &    \times
        \sin^2(a_q^{(2)}\tau) \}.
\end{eqnarray}
The magnetization (or spin-polarization) is defined as
$P_z=(n_{\uparrow}-n_{\downarrow})/(n_{\uparrow}+n_{\downarrow})$.
We have
\begin{eqnarray}
 P_z
 &=& \cos(2\rho_q^{(1)})
    +[ \cos(4\rho_q^{(2)}-2\rho_q^{(1)})
      -\cos(2\rho_q^{(1)}) ] \nonumber \\
 & &    \times
     \sin^2(a_q^{(2)}\tau).
\end{eqnarray}
This formula gives a clear picture of a harmonic
$\theta$-independent oscillation with a period
$\tau_p=\pi/a_q^{(2)}$ and an amplitude $A_{\mathrm{amp}}
=|\cos(4\rho_q^{(2)}-2\rho_q^{(1)})-\cos(2\rho_q^{(1)})|$.

Based on the second set of parameters let us define the energy
difference of the two components as $E_{\mathrm{diff}}^{(2)}
\equiv\frac{1}{2}|\langle\varphi_{q-\beta,\uparrow}|
\hat{h}^{(2)}|\varphi_{q-\beta,\uparrow}\rangle
-\langle\varphi_{q+\beta,\downarrow}| \hat{h}^{(2)}|
\varphi_{q+\beta,\downarrow}\rangle|=(2q\beta-\delta_2)$. Then,
the frequency of oscillation $1/\tau_p
=\frac{1}{\pi}\sqrt{(E_{\mathrm{diff}}^{(2)})^2+\gamma_2^2}$.
This formula demonstrates that both the strength of the Raman
coupling and the energy difference are crucial. A larger
$|\gamma_2|$ leads always to a higher frequency
$\geq|\gamma_2|/\pi$. This frequency can be further tuned by
varying $\delta_2$ around $2q\beta$. In particular, when
$\delta_2$ is tuned so that $E_{\mathrm{diff}}^{(2)}$ is zero,
$1/\tau_p$ will arrive at its conditional minimum
$|\gamma_2|/\pi$. Since $E_{\mathrm{diff}}^{(2)}$ depends on
$q$ but not on the other details of the initial state,
$1/\tau_p$ depends essentially on the second set of parameters
when $q$ is given.

On the other hand, the amplitude $A_{\mathrm{amp}}$ depends on
both sets of parameters explicitly. When the first set of
parameter has been given so that $\rho_q^{(1)}$ is fixed,
$A_{\mathrm{amp}}$ will arrive at its conditional maximum at a
specific value of $\rho_q^{(2)}$, namely,
$\rho_q^{(2)}=\rho_q^{(1)}/2$ (if $\gamma_2<0$ and
$\rho_q^{(1)}<-\pi/4$), or $=\rho_q^{(1)}/2+\pi/2$ (if
$\gamma_2>0$ and $\rho_q^{(1)}<-\pi/4$), or
$=\rho_q^{(1)}/2-\pi/4$ (if $\gamma_2<0$ and
$-\pi/4\leq\rho_q^{(1)}\leq\pi/4$), or $=\rho_q^{(1)}/2+\pi/4$
(if $\gamma_2>0$ and $-\pi/4\leq\rho_q^{(1)}\leq\pi/4$), or
$=\rho_q^{(1)}/2-\pi/2$ (if $\gamma_2<0$ and
$\rho_q^{(1)}>\pi/4$), or $=\rho_q^{(1)}/2$ (if $\gamma_2>0$
and $\rho_q^{(1)}>\pi/4$). It implies that, no matter how the
first set of parameters are, one can tune the second set so
that the amplitude is conditionally maximized, namely,
$A_{\mathrm{amp}}=1+|\cos(2\rho_q^{(1)})|$. In particular, when
$\gamma_1=0 $, we have $\rho_q^{(1)}=\pm\pi/2$ or 0. In this
case, if $\gamma_2\neq 0$ and $\delta_2$ is tuned so that
$E_{\mathrm{diff}}^{(2)}=0$, then
$\rho_q^{(2)}=(\gamma_2/|\gamma_2|)\pi/4$ and the amplitude
would arrive at its absolute maximum, namely,
$A_{\mathrm{amp}}=2$. Note that the condition
$\rho_q^{(1)}=\pm\pi/2$ or 0 implies that the initial state is
pure without mixing (namely, either
$\psi_{\mathrm{init}}=\varphi_{q-\beta,\uparrow}$ or
$\varphi_{q+\beta,\downarrow}$), while the fact
$A_{\mathrm{amp}}=2$ implies a complete transformation of the
two components (namely, from a pure up-state to a pure
down-state, or vice versa). Therefore, $\gamma_1=0$ is required
so that the maximal amplitude could be realized. On the other
hand, in any cases, when $\gamma_2$ and $\gamma_1$ have the
same sign and $\rho_q^{(2)}\rightarrow\rho_q^{(1)}$, or
$\gamma_2$ and $\gamma_1$ have opposite signs and
$\rho_q^{(2)}\rightarrow\rho_q^{(1)}\pm\pi/2$,
$A_{\mathrm{amp}}\rightarrow 0$ and the oscillation damps. This
is obvious from the expression of $A_{\mathrm{amp}}$.

In addition to the densities, one can further define the
current based on the equation of continuity as
$-\frac{\partial}{\partial t}(\psi^+(\theta,t)\psi(\theta,t))
=\frac{\partial\ j}{\partial\theta}$. From this definition we
found that the current $j=j_{\uparrow}+j_{\downarrow}$ in which
$j_{\uparrow}\equiv j_{\mathrm{unit}}n_{\uparrow}(q-\beta)$ is
the current of the up-component and is named up-current, while
$j_{\downarrow}\equiv j_{\mathrm{unit}}n_{\downarrow}(q+\beta)$
is the down-current, where the unit is
$j_{\mathrm{unit}}=\hbar/(mR^2)$.\cite{toz} The currents are
also oscillating with the same period $\pi/a_q^{(2)}$, and they
are also $\theta$-independent. Obviously, when $|\beta|>|q|$,
$j_{\uparrow}$ and $j_{\downarrow}$ will have different signs.
Therefore, counter propagating currents emerge. This is a
distinguished feature.

The above formulae arise from a single-particle Hamiltonian.
For N-particle systems, when all the particles fall into the
same state $\psi_{\mathrm{gs}}$ initially and the interaction
is neglected, it is straight forward to prove that the above
formulae hold also.

To give numerical results, the radius is given at $R=12\mu m$
in this paper. A contour diagram for the frequency $1/\tau_p$
versus $\gamma_2$ and $\delta_2$ is shown in Fig.1. Where the
minimum is located at $\gamma_2=0$ and
$E_{\mathrm{diff}}^{(2)}=0$ (namely, $\delta_2=2q\beta=-18)$,
at which the oscillation vanishes.

\begin{figure}[tbp]
 \begin{center}
 \resizebox{0.95\columnwidth}{!}{ \includegraphics{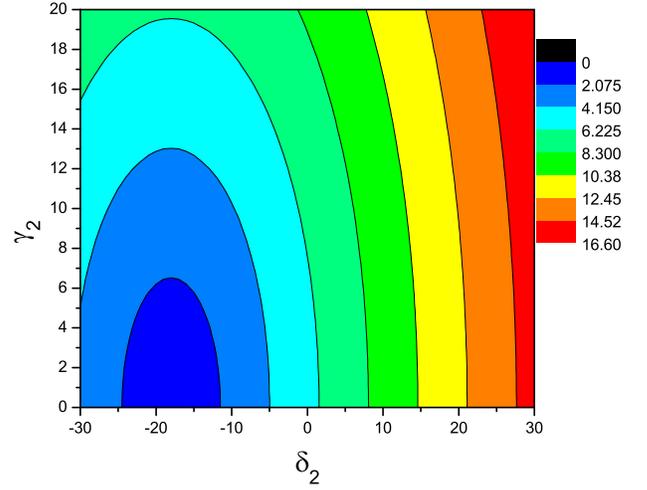} }
 \caption{The frequency $1/\tau_p$ versus $\delta_2$ and
$\gamma_2$. The oscillation is caused by a sudden change in
$\gamma$ and $\delta$. $\beta=3$ and $q=-3$ are given. The
values of the contours are in an arithmetic series. The contour
closest to the right side has the largest value
$1/\tau_p=14.5$.}
 \end{center}
\end{figure}

Two contour diagrams for $A_{\mathrm{amp}}$ versus $\gamma_2$
and $\delta_2$ with $\gamma_1$ and $\delta_1$ being fixed are
shown in Fig.2, where all the contours appear as straight
lines. This arises from the fact that $\rho_q^{(2)}$ can be
rewritten as a function of
$|\gamma_2/E_{\mathrm{diff}}^{(2)}|=|\gamma_2/(\delta_2-2q\beta)|$
together with the two signs of $\gamma_2$ and
$\delta_2-2q\beta$ (it implies that a given $\rho_q^{(2)}$ is
associated with a straight line on the $\delta_2-\gamma_2$
plane). In Fig.2a (b) the given values of $\beta$, $\gamma_1$,
and $\delta_1$ lead to $q=-3\ (3)$ and $\rho_q^{(1)}=0.435\pi \
(0.0653\pi)$. According to the discussion given previously and
when $\gamma_2>0$, the maximum of $A_{\mathrm{amp}}$ should
appear when $\rho_q^{(2)}=\rho_q^{(1)}/2=0.218\pi$ in Fig.2a,
and $\rho_q^{(2)}=\rho_q^{(1)}/2+\pi/4=0.315\pi$ in 2b. This
maximum is associated with the straight dotted line marked in
the figure (Incidentally, if $\rho_q^{(2)}=\pi/4$, then the
associated straight line would be vertical. Since the two
dotted lines have their $\rho_q^{(2)}$ close to $\pi/4$, they
are nearly vertical). Whereas when $\rho_q^{(2)}=\rho_q^{(1)}$,
the associated straight line is marked by a solid line in the
figure (where the small circle marks the place that
$\gamma_2=\gamma_1$, and $\delta_2=\delta_1$). Three notable
features are reminded:

(i) The frequency does not depend on the first set of
parameters, except $q$. A larger $|\gamma_2|$ leads always to a
higher frequency which can be further tuned by varying
$\delta_2$ around $2q\beta$. When $\delta_2=2q\beta$, the
frequency arrives at its conditional minimum $|\gamma_2|/\pi$.

(ii) $A_{\mathrm{amp}}$ depends on the ratios
$\gamma_1/\delta_1$ and $\gamma_2/\delta_2$. When $\gamma_2$
and $\delta_2$ are given along the solid line (the slope of
this line depends on $\gamma_1/\delta_1$), the sudden change of
$\gamma_1\rightarrow \gamma_2$ and $
\delta_1\rightarrow\delta_2$ can not cause an evolution (i.e.,
$A_{\mathrm{amp}}=0$). When $\gamma_2$ and $\delta_2$ are given
along the dotted line (also depends on $\gamma_1/\delta_1$),
the amplitude will remain unchanged and is conditionally
maximized, namely, $A_{\mathrm{amp}}=1+|\cos(2\rho_q^{(1)})|$.
In any cases, \textit{one can always tune }$ \delta_2$
\textit{\ so that the amplitude is either suppressed or
conditionally maximized}.

(iii) Due to the fact that the contours are straight lines,
when $\gamma_2$ is small ($\neq 0$) there is a narrow domain of
$\delta_2$ surrounding $2q\beta$ in which $A_{\mathrm{amp}}$ is
highly sensitive to $ \delta_2$ and appears as a sharp peak. At
the peak the amplitude is large while the associated frequency
is low.

\begin{figure}[tbp]
 \begin{center}
 \resizebox{0.95\columnwidth}{!}{ \includegraphics{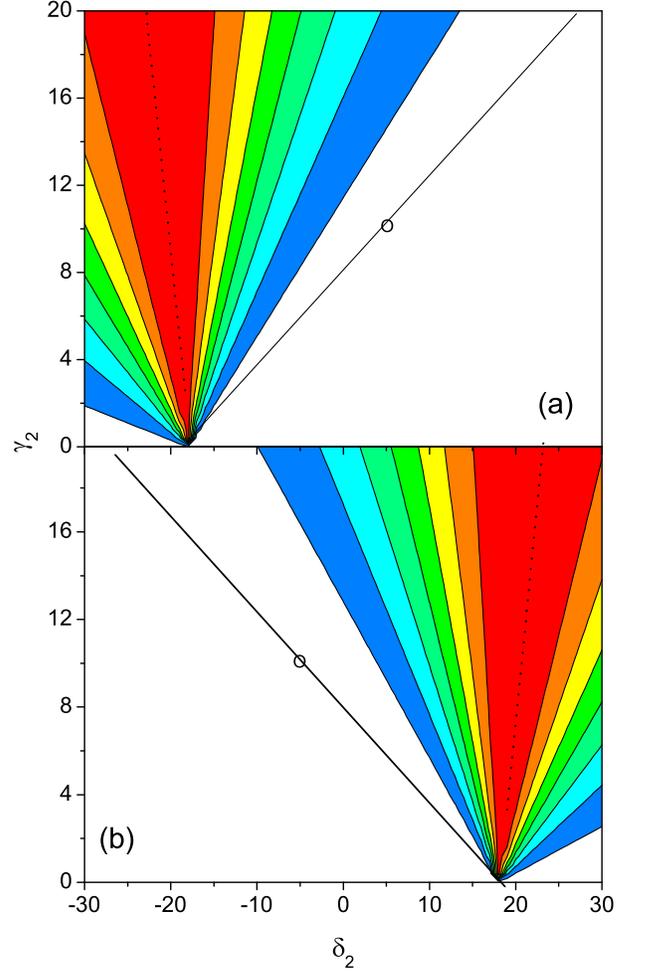} }
 \caption{The amplitude $A_{\mathrm{amp}}$ versus $\delta_2$ and
$\gamma_2$. $\beta=3$, $\gamma_1=10$, and $\delta_1=5(-5)$, are
given in a (b). Accordingly, $q=-3\ (3)$ in a (b). The
oscillation is caused by a sudden change from $\gamma_1$ to
$\gamma_2$ and $\delta_1$ to $\delta_2$. The dotted line marks
the locations where $\rho_q^{(2)}=\rho_q^{(1)}/2$ (a), or
$\rho_q^{(2)}=\rho_q^{(1)}/2+\pi/4$ (b). Thus this line marks
the conditional maximum of $A_{\mathrm{amp}}
=1+|\cos(2\rho_q^{(1)})|=1.917$ in both (a) and (b), which is
the largest value in each panel. The contours going away from
this line are in an arithmetic series and decrease to zero. The
black small circle marks the location where $\delta_2=\delta_1$
and $\gamma_2=\gamma_1$. The solid line passing through the
circle marks the locations where $\rho_q^{(2)}=\rho_q^{(1)}$
and accordingly $A_{\mathrm{amp}}=0$.}
 \end{center}
\end{figure}

\section{Evolution initiated from a superposition state}

Recently, the initial state was prepared in a superposition
state in an experiment of cold atoms with the ring
geometry.\cite{sb} In this experiment an additional radio
frequency field was used for preparing the initial state
\begin{equation}
 \psi_{\mathrm{init}}
  =  \sin(\phi/2)
     \varphi_{q\uparrow}
    +\cos(\phi/2)
     \varphi_{q\downarrow},
\end{equation}
where $\phi$ determines the initial ratio of the two components
and is tunable ($0\leq\phi\leq\pi$). $q$ is a tunable integer
and marks the initial angular momentum of the particle. In this
case we have the set of parameters $\phi$, $q$, $\beta$,
$\gamma$, and $\delta$. The corresponding time-dependent
solution is
\begin{eqnarray}
 \psi(\theta,t)
 &=& e^{-i\tau(q^2+2\beta^2)}
     ( f_{u_1}
       \varphi_{q,\uparrow}
      +f_{u_2}
       \varphi_{q-2\beta,\uparrow} \nonumber \\
 & &  +f_{d_1}
       \varphi_{q+2\beta,\downarrow}
      +f_{d_2}
       \varphi_{q,\downarrow}),
\end{eqnarray}
where $f_{u_1} =\sin(\phi/2)e^{-2iq\beta\tau}
[\cos(a_{q+\beta}\tau)+i\cos(2\rho_{q+\beta})\sin(a_{q+\beta}\tau)]$,
$f_{u_2} =-i\cos(\phi/2)e^{2iq\beta\tau}
\sin(2\rho_{q-\beta})\sin(a_{q-\beta}\tau)$ , $f_{d_1}
=-i\sin(\phi/2)e^{-2iq\beta\tau}
\sin(2\rho_{q+\beta})\sin(a_{q+\beta}\tau)$, and $f_{d_2}
=\cos(\phi/2)e^{2iq\beta\tau}
[\cos(a_{q-\beta}\tau)-i\cos(2\rho_{q-\beta})\sin(a_{q-\beta}\tau)]$.

Accordingly, we can obtain the up- and down-densities
$n_{\uparrow}(\theta,\tau)$ and $n_{\uparrow}(\theta,\tau)$
given in the appendix. From them the magnetization $P_z$ can be
obtained. It is apparent that, when $\phi=\pi$ or 0 (i.e.,
$\psi_{\mathrm{init}}$ is a pure up-state or down-state) the
magnetization has a very simple form as
\begin{equation}
 P_z
  =  \cos(4\rho_{q+\beta})
    +( 1-\cos(4\rho_{q+\beta}))
     \cos^2(a_{q+\beta}\tau),
\end{equation}
if $\phi=\pi$, or
\begin{equation}
 P_z
  = -\cos(4\rho_{q-\beta})
    -( 1-\cos(4\rho_{q-\beta}))
     \cos^2(a_{q-\beta}\tau),
\end{equation}
if $\phi=0$.

These formulae also give a clear picture of harmonic
oscillation, but the periods are different for the two cases of
$\phi$.

When $\phi=\pi$, $\tau_p=\pi/a_{q+\beta}$. If $\delta$ is tuned
so that $\delta=2(q+\beta)\beta\equiv\delta_o$, the frequency
will be minimized, namely,
$(1/\tau_p)_{\min}=\frac{1}{\pi}|\gamma|$. When $\delta$ goes
away from $\delta_o$, the frequency increases. The amplitude
$A_{\mathrm{amp}}=1-\cos(4\rho_{q+\beta})$. Obviously, when
$\delta=\delta_o$, we have $\rho_{q+\beta}=\pm\pi/4$, and the
amplitude would arrive at its maximum, namely,
$(A_{\mathrm{amp}})_{\max}=2$ . Further more, when
$\gamma\rightarrow\pm 0$, one can prove that $\rho_{q+\beta}$
will tend to 0 or $\pm\pi/2$. In any of these cases
$A_{\mathrm{amp}}=0$ and the oscillation disappear. This
implies that $\gamma$ is the source of the oscillation.

When $\phi=0$, the discussion in the preceding paragraph holds
also, except that $q+\beta $ should be changed to $q-\beta$,
and $\delta_o$ should be replaced by $2(q-\beta)\beta$. An
example of $A_{\mathrm{amp}}$ is shown in Fig.3. Fig.3 is of a
similar form to Fig.2 (all the contours are straight lines
converge to a point with $\gamma=0$) except that the dotted
line in the former is exactly vertical.

\begin{figure}[tbp]
 \begin{center}
 \resizebox{0.95\columnwidth}{!}{ \includegraphics{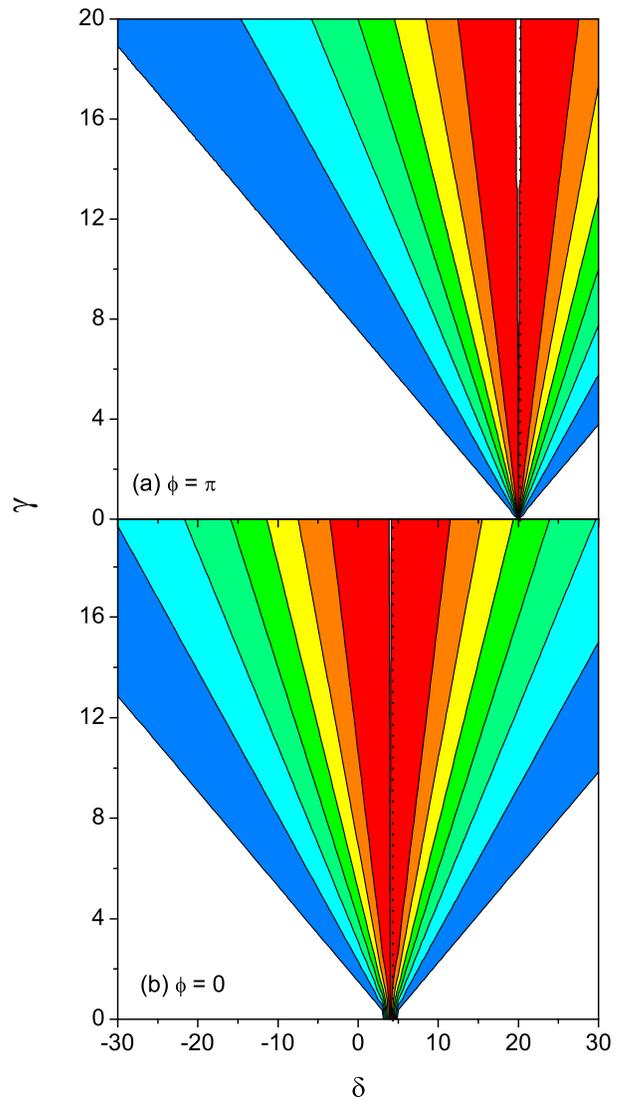} }
 \caption{The amplitude $A_{\mathrm{amp}}$ versus $\delta$ and
$\gamma$ for the case that the initial state is a superposition
state. $q=3$ and $\beta=2$ are assumed. $\phi=\pi(0)$ are given
in a (b). The dotted line marks the maximum of
$A_{\mathrm{amp}}=2 $. The locations with $\gamma=0$ have
$A_{\mathrm{amp}}=0$.}
 \end{center}
\end{figure}

When $\phi$ is neither $\pi$ nor 0, the oscillation is no more
harmonic. In particular, both the up- and down-densities depend
on $\theta$ (refer to the appendix). It implies that the
densities are not uniform on the ring as before. Consequently,
stripes emerge along the ring, and the stripes move with time.
An example of the evolution of $P_z$ is shown in Fig.4. Where
the relation $\delta=2(q+\beta)\beta$ holds. Accordingly, the
curve with $\phi=\pi$ has the maximized $A_{\mathrm{amp}}=2$
and a minimized frequency. Whereas the curve with $\phi=0$ has
a smaller $A_{\mathrm{amp}}$ and a larger frequency. For
$\phi=\pi/2$, the anharmonic oscillation is clearly shown.

\begin{figure}[tbp]
 \begin{center}
 \resizebox{0.95\columnwidth}{!}{ \includegraphics{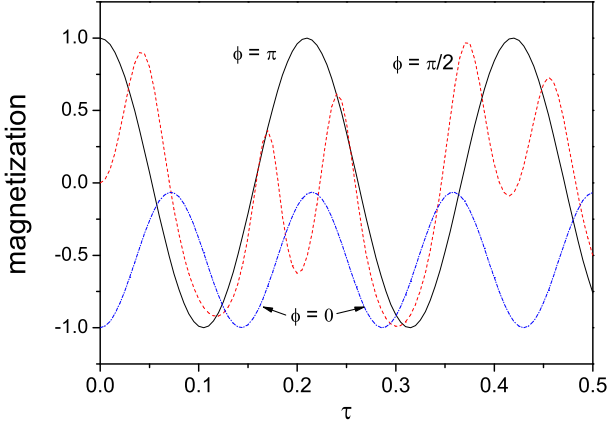} }
 \caption{The magnetization $P_z$ versus $\tau$ observed at
$\theta=0$ for the case that the initial state is a
superposition state with a $\phi$ marked by the curve. $q=3$,
$\beta=2$, $\gamma=15$, and $\delta=20$ are assumed. When
$\phi=\pi$ or 0 the oscillation is harmonic. Otherwise, it is
not.}
 \end{center}
\end{figure}

An example of the up-density versus $\theta$ with $\phi=\pi/2$
is shown in Fig.5, where the stripes emerge along the ring.
Note that two waves $\varphi_{q,\uparrow}$ and
$\varphi_{q-2\beta,\uparrow}$ are contained in the
up-component, and $\varphi_{q,\downarrow}$ and
$\varphi_{q+2\beta,\downarrow}$ in the down-component. The
stripes arise from the interference of these waves. Note that
one of these waves contains an additional factor $e^{\mp
i2\beta\theta}$. Accordingly, the stripes appear periodically
with $\theta$ as shown in Fig.5 and the number of peaks
(valleys) in the domain $(0,2\pi)$ is $2\beta$ (say, the number
is four in Fig.5, where $\beta=2$). The locations of the peaks
of the stripes move with time, the magnitudes of the peaks also
change with time.

As before, when all the particles are given in the same
$\psi_{\mathrm{init}}$ initially and the interaction is
neglected, the above formulae and discussion hold also for
N-particle systems.

\begin{figure}[tbp]
 \begin{center}
 \resizebox{0.95\columnwidth}{!}{ \includegraphics{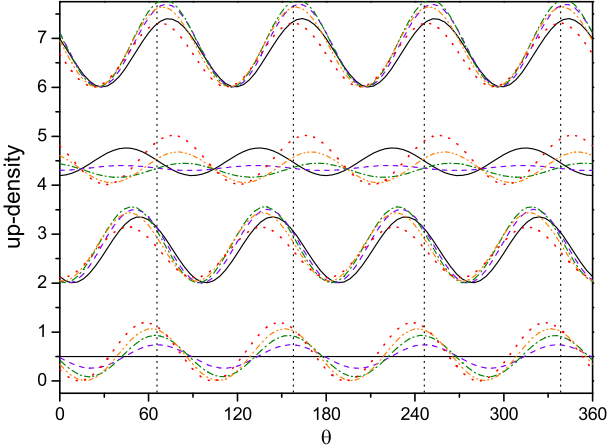} }
 \caption{Up-density $n_{\uparrow}$ versus $\theta$ with
$\phi=\pi/2$. The densities are given by 20 curves for a
sequence of $\tau$. Starting from $\tau=0$, in each step $\tau$
is increased by $1/40$. The 20 curves in the sequence are
divided into 4 groups. The $n_{\uparrow}$ of each group has
been shifted up by 2 relative to its preceding group, for a
clearer perspective. The five $n_{\uparrow}$ in each group are
plotted in solid, dash, dash-dot, dash-dot-dot, and dot lines,
respectively, according to the time-sequence ($\tau=0$ is the
horizontal solid line in the lowest group). The parameters are
$q=1$, $ \beta=2$, $\gamma=10$, and $\delta=1$. The vertical
dotted lines are also for guiding the eyes.}
 \end{center}
\end{figure}

\section{Evaluation of the effect of interaction}

When the interaction is taken into account, the total
Hamiltonian is $H=\sum_i\hat{h}_i+\sum_{i<j}V_{ij}$, where
$V_{ij}=g\delta(\theta_i-\theta_j)$, $g=g_{\upuparrows}$ (if
both atoms are up), $g_{\downdownarrows}$ (both down), or
$g_{\uparrow\downarrow}$ (one up and one down). In order to
evaluate the effect of interaction in the simplest way, we
study a two-body system. Firstly, a set of single particle
states
$\varphi_{k\mu}=\frac{1}{\sqrt{2\pi}}e^{ik\theta}\chi_{\mu}$,
where $\chi_{\mu}=\binom{1}{0}$ or $\binom{0}{1}$, and
$-k_{\max}\leq k\leq k_{\max}$, are adopted. Accordingly, we
have $2(2k_{\max}+1)$ single particle states, and they are
renamed as $\varphi_j\equiv\varphi_{k_j\mu_j}$. Based on
$\varphi_j$ a set of basis functions for the two-body system
$\Phi_i=\tilde{S}[\varphi_j(1)\varphi_{j'}(2)]$ are defined,
where $\tilde{S}$ is for the symmetrization and normalization,
and $j\geq j'$.

Note that the strength of a pair of realistic Rb atoms is
$g_{\mathrm{Rb}}=7.79\times 10^{-12}H_Z cm^3$ (the differences
in the strengths between the up-up, up-down, and down-down
pairs are very small and are therefore neglected). Since the
atoms in related experiments are not distributed exactly on a
one-dimensional ring but in a domain surrounding the ring, the
effect of the diffused distribution should be considered.
Hence, for each $\varphi_{k\mu}$ we define its 3-dimensional
counterpart $\varphi_{k\mu}^{[3d]}=\frac{1}{\pi
r_w}\sqrt{\frac{1}{R}}e^{ik\theta}e^{-(r/r_w)^2}\chi_{\mu}$,
where the Gaussian function $e^{-(r/r_w)^2}$ describes the
diffused distribution and $r_w$ measures the width of the
distribution. With $ \varphi_{k\mu}^{[3d]}$ we define an
effective strength $g_{\mathrm{eff}}$ so that for any pair of
matrix elements
\begin{eqnarray}
 & & g_{\mathrm{eff}}
     \int d\theta_i d\theta_j
     \varphi_{k_1\mu_1}^{\dagger}
     \varphi_{k_2\mu_2}^{\dagger}
     \delta(\theta_i-\theta_j)
     \varphi_{k_3\mu_3}
     \varphi_{k_4\mu_4} \nonumber \\
 &=& g_{\mathrm{Rb}}
     \int d\mathbf{r}_i d\mathbf{r}_j
     \varphi_{k_1\mu_1}^{[3d]\dagger}
     \varphi_{k_2\mu_2}^{[3d]\dagger}
     \delta(\mathbf{r}_i-\mathbf{r}_j)
     \varphi_{k_3\mu_3}^{[3d]}
     \varphi_{k_4\mu_4}^{[3d]}. \ \ \
\end{eqnarray}
Then, we have $g_{\mathrm{eff}}=\frac{1}{\pi R\
r_w^2}g_{\mathrm{Rb}}$ which is the effective strength adopted
in our calculation.

\begin{figure}[tbp]
 \begin{center}
 \resizebox{0.95\columnwidth}{!}{ \includegraphics{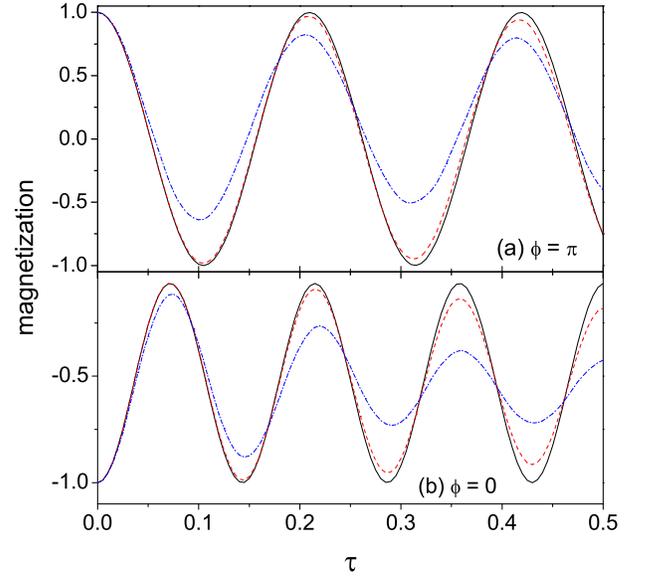} }
 \caption{$P_z$ of a 2-body system of Rb atoms versus $\tau$
with interaction $V_{ij}=g\delta(\theta_i-\theta_j)$, $g=0$
(solid), $100g_{\mathrm{eff}}$ (dash), and
$2000g_{\mathrm{eff}}$ (dash-dot). The initial state has
$\phi=\pi$ (a) and $=0$ (b). $r_w=R/\sqrt{18}$. The other
parameters are the same as in Fig.4 }
 \end{center}
\end{figure}

It is assumed that both atoms are given at the same
superposition state with $\phi=\pi$ or 0 initially. When the
Hamiltonian is diagonalized in the space expanded by $\Phi_i$,
the eigenenergies $E_{l}$ and eigenstates $\Psi_l\equiv
\sum_iC_{li}\Phi_i$ can be obtained, and the time-dependent
state is
\begin{equation}
 \Psi(\theta,t)
  =  \sum_l
     e^{-i\tau E_l}
     \Psi_l\langle
     \Psi_l|
     \Psi_{\mathrm{init}}\rangle,
\end{equation}
where $\Psi_{\mathrm{init}}
=\psi_{\mathrm{init}}(1)\psi_{\mathrm{init}}(2) $. From
$\Psi(\theta,t)$ the densities and $P_z(t)$ can be calculated.
An examples with $k_{\max}=14$ is shown in Fig.6 where the
strength $g$\ is given at three values. When $k_{\max}$ is
changed from 14 to 12, there is no explicit changes in the
pattern. It implies that the choice $k_{\max}=14$ is sufficient
in qualitative sense.

In Fig.6 the curves "1" and "2" overlap approximately. It
implies that the effect of interaction with a $g\leq
100g_{\mathrm{eff}}$ is very small. However, when
$g=2000g_{\mathrm{eff}}$, the amplitude decreases with time
explicitly as shown by the dash-dot curve, while the period
remains nearly unchanged. Note that, based on the
Gross-Pitaevskii equation, the effect of the combined
interaction imposing on a single particle from the other $N-1$
particles is similar to an appropriate strengthening of the
strength of the atom-atom interaction. Thus the explicit damp
that happens when $g=2000g_{\mathrm{eff}}$ implies that, for an
N-body system with a larger $N$, the damp would actually occur.
This is a topic to be clarified further. In fact, the damp of
the oscillation has already been observed in all related
existing experiments.\cite{jyz,yjl1}

\section{Summary}

In summary, the oscillation of the cold atoms under the SOC and
constrained on a ring has been studied analytically without
taking the interaction into account. Then, the effect of the
interaction is evaluated numerically via a two-body system. Two
cases, namely, the evolution that starts from a g.s. and is
induced by a sudden change of the laser field, and the
evolution that starts from a superposition state, are involved.
The emphasis is placed on clarifying the relation between the
parameters of the laser beams and the period and amplitude of
the oscillation. This is achieved by giving a set of formulae
so that the relation can be understood analytically. The
conditions that the frequency and/or the amplitude can be
maximized or minimized have been predicted. To realize the
mutual-transformation of the two components, not only the
strength of the Raman coupling $\gamma$ is important, the
energy difference between the two components is also important
which can be tuned by $\delta$ (this is similar to the case of
a resonance). The frequency of oscillation will increase with
$\gamma$ in general, while the amplitude depends on the
interrelation of $\gamma$ and $\delta$. When $(\delta,\gamma)$
vary along a specific straight line on the $\delta-\gamma$
plane, the amplitude may increase, decrease, or even remain
unchanged depending on the slope of the line. This is a common
feature for the first case and for the second case with
$\phi=\pi$ or $0$. Whereas when $\phi\neq\pi$ and $0$, movable
stripes will emerge on the ring. Experimental confirmation of
the regularity unveiled in this paper and the predicted
phenomena is expected.

The support from the NSFC (China) under the grant number
10874249 is appreciated.

\section*{Appendix}

When the initial state is $\psi_{\mathrm{init}}
=\sin(\phi/2)\varphi_{q\uparrow}+\cos(\phi/2)\varphi_{q\downarrow}$,
the associated up- and down-densities during the evolution are

\begin{widetext}
\begin{eqnarray}
 n_{\uparrow}(\theta,\tau)
 &=& \frac{1}{2\pi}
     \{ \sin^2(\phi/2)
        [ \cos^2(a_{q+\beta}\tau)
         +\cos^2(2\rho_{q+\beta})
          \sin^2(a_{q+\beta}\tau )] \nonumber \\
 & &   +\cos^2(\phi/2)
        \sin^2(2\rho_{q-\beta})
        \sin^2(a_{q-\beta}\tau) \nonumber \\
 & &   -\sin\phi
        \sin(2\rho_{q-\beta})
        \sin(a_{q-\beta}\tau)
        [ \cos(2\beta\theta-4q\beta\tau)
          \cos(2\rho_{q+\beta})
          \sin(a_{q+\beta}\tau) \nonumber \\
 & &     +\sin(2\beta\theta-4q\beta\tau)
          \cos(a_{q+\beta}\tau)] \},
\end{eqnarray}
and
\begin{eqnarray}
 n_{\downarrow}(\theta,\tau)
 &=& \frac{1}{2\pi}
     \{ \cos^2(\phi/2)
        [ \cos^2(a_{q-\beta}\tau)
         +\cos^2(2\rho_{q-\beta})
          \sin^2(a_{q-\beta}\tau) ] \nonumber \\
 & &   +\sin^2(\phi/2)
        \sin^2(2\rho_{q+\beta})
        \sin^2(a_{q+\beta}\tau) \nonumber \\
 & &   +\sin\phi
        \sin(2\rho_{q+\beta})
        \sin(a_{q+\beta}\tau)
        [ \cos(2\beta\theta-4q\beta\tau)
          \cos(2\rho_{q-\beta})
          \sin(a_{q-\beta}\tau) \nonumber \\
 & &     +\sin(2\beta\theta-4q\beta\tau)
          \cos(a_{q-\beta}\tau) ] \}.
\end{eqnarray}
\end{widetext}

\end{document}